\documentclass[preprint,tighten]{aastex}
\usepackage{lineno}

\def\obslivetime{18.6 hours}
\def\surveyefflivetime{$\sim 30$ hours} 

\def\sizepe{70}

\def\mslrange{$\rm 0.05<MSL<1.25$}
\def\mswrange{$\rm 0.05<MSW<1.10$}

\def\analysisthreshold{320~GeV}

\def\posttrials{7.5 standard deviations}
\def\posttrialssig{$7.5\sigma$}
\def\excesscounts{$319 \pm 39$}
\def\specradius{$0.24^{\circ}$}

\def\containment{$0.09^{\circ}$}

\def\fittedcentroidsexg{RA $20^{h} 20^{m} 04.8^{s}$, Dec $\mathrm +40^{\circ} 45^{\prime} 36^{\prime\prime}$ (J2000)}
\def\fittedextension{$\rm 0.23^{\circ} \pm {0.03^{\circ}_{\mathrm stat}} {}^{+0.04^{\circ}}_{-0.02^{\circ}{\mathrm sys}}$}

\def\energyres{15\%}
\def\threshold{320~GeV}

\def\spechisq{$\rm \chi^{2}/d.o.f = 8.76/6$}
\def\specindexg{$\rm \Gamma = 2.37 \pm {0.14}_{\mathrm stat} \pm {0.20}_{\mathrm sys}$}
\def\specindex{$\rm 2.37 \pm {0.14}_{\mathrm stat} \pm {0.20}_{\mathrm sys}$}
\def\fluxnorm{$\rm N_{0} = 1.5 \pm 0.2_{stat} \pm {0.4}_{sys} \times 10^{-12} \ ph  \ {TeV}^{-1} \ cm^{-2} \ s^{-1}$}
\def\integralflux{$\rm 5.2 \pm 0.8 _{stat} \pm {1.4}_{sys} \times 10^{-12} \ ph \ cm^{-2} \ s^{-1}$}

\def\fluxpercentage{3.7\%} 

\def\initialdensity{$0.5$~cm$^{-3}$}
\def\inferredensities{$1.0-5.5$~cm$^{-3}$}

\def\centroidpulsaroffset{$\sim 0.5^{\circ}$}

\def\snr{SNR~G78.2+2.1}
\def\ver{VER~J2019+407}

\def\fgl{1FGL~J2020.0+4049}
\def\fgltwo{2FGL~J2019.1+4040}
\def\fglpsr{1FGL~J2021.5+4026}
\def\psr{PSR~J2021+4026}

\def\rosat{{\it ROSAT}}

\newcommand\veritas{{VERITAS}}

\newcommand\fermi{{\it Fermi}}

\slugcomment{Accepted by the Astrophysical Journal}

\shorttitle{TeV gamma-ray emission from SNR~G78.2+2.1}
\shortauthors{Aliu et al.}

\begin{document}

\title{Discovery of TeV Gamma-ray Emission Toward \\ Supernova Remnant SNR~G78.2+2.1}

\author{
E.~Aliu\altaffilmark{1},
S.~Archambault\altaffilmark{2},
T.~Arlen\altaffilmark{3},
T.~Aune\altaffilmark{3},
M.~Beilicke\altaffilmark{4},
W.~Benbow\altaffilmark{5},
R.~Bird\altaffilmark{6},
A.~Bouvier\altaffilmark{7},
S.~M.~Bradbury\altaffilmark{8},
J.~H.~Buckley\altaffilmark{4},
V.~Bugaev\altaffilmark{4},
K.~Byrum\altaffilmark{9},
A.~Cannon\altaffilmark{6},
A.~Cesarini\altaffilmark{10},
L.~Ciupik\altaffilmark{11},
E.~Collins-Hughes\altaffilmark{6},
M.~P.~Connolly\altaffilmark{10},
W.~Cui\altaffilmark{12},
R.~Dickherber\altaffilmark{4},
C.~Duke\altaffilmark{13},
J.~Dumm\altaffilmark{14},
V.~V.~Dwarkadas\altaffilmark{15},
M.~Errando\altaffilmark{1},
A.~Falcone\altaffilmark{16},
S.~Federici\altaffilmark{17,18},
Q.~Feng\altaffilmark{12},
J.~P.~Finley\altaffilmark{12},
G.~Finnegan\altaffilmark{19},
L.~Fortson\altaffilmark{14},
A.~Furniss\altaffilmark{7},
N.~Galante\altaffilmark{5},
D.~Gall\altaffilmark{20},
G.~H.~Gillanders\altaffilmark{10},
S.~Godambe\altaffilmark{19},
E.~V.~Gotthelf\altaffilmark{21},
S.~Griffin\altaffilmark{2},
J.~Grube\altaffilmark{11},
G.~Gyuk\altaffilmark{11},
D.~Hanna\altaffilmark{2},
J.~Holder\altaffilmark{22},
H.~Huan\altaffilmark{23},
G.~Hughes\altaffilmark{17},
T.~B.~Humensky\altaffilmark{24},
P.~Kaaret\altaffilmark{20},
N.~Karlsson\altaffilmark{14},
M.~Kertzman\altaffilmark{25},
Y.~Khassen\altaffilmark{6},
D.~Kieda\altaffilmark{19},
H.~Krawczynski\altaffilmark{4},
F.~Krennrich\altaffilmark{26},
M.~J.~Lang\altaffilmark{10},
K.~Lee\altaffilmark{4},
A.~S~Madhavan\altaffilmark{26},
G.~Maier\altaffilmark{17},
P.~Majumdar\altaffilmark{3,27},
S.~McArthur\altaffilmark{23},
A.~McCann\altaffilmark{28},
J.~Millis\altaffilmark{29},
P.~Moriarty\altaffilmark{30},
R.~Mukherjee\altaffilmark{1},
T.~Nelson\altaffilmark{14},
A.~O'Faol\'{a}in de Bhr\'{o}ithe\altaffilmark{6},
R.~A.~Ong\altaffilmark{3},
M.~Orr\altaffilmark{26},
A.~N.~Otte\altaffilmark{31},
D.~Pandel\altaffilmark{32},
N.~Park\altaffilmark{23},
J.~S.~Perkins\altaffilmark{33,34},
M.~Pohl\altaffilmark{18,17},
A.~Popkow\altaffilmark{3},
H.~Prokoph\altaffilmark{17},
J.~Quinn\altaffilmark{6},
K.~Ragan\altaffilmark{2},
L.~C.~Reyes\altaffilmark{35},
P.~T.~Reynolds\altaffilmark{36},
E.~Roache\altaffilmark{5},
H.~J.~Rose\altaffilmark{8},
J.~Ruppel\altaffilmark{18,17},
D.~B.~Saxon\altaffilmark{22},
M.~Schroedter\altaffilmark{5},
G.~H.~Sembroski\altaffilmark{12},
G.~D.~\c{S}ent\"{u}rk\altaffilmark{24},
C.~Skole\altaffilmark{17},
I.~Telezhinsky\altaffilmark{18,17},
G.~Te\v{s}i\'{c}\altaffilmark{2},
M.~Theiling\altaffilmark{12},
S.~Thibadeau\altaffilmark{4},
K.~Tsurusaki\altaffilmark{20},
J.~Tyler\altaffilmark{2},
A.~Varlotta\altaffilmark{12},
V.~V.~Vassiliev\altaffilmark{3},
S.~Vincent\altaffilmark{17},
S.~P.~Wakely\altaffilmark{23},
J.~E.~Ward\altaffilmark{4},
T.~C.~Weekes\altaffilmark{5},
A.~Weinstein\altaffilmark{26,*},
T.~Weisgarber\altaffilmark{23},
R.~Welsing\altaffilmark{17},
D.~A.~Williams\altaffilmark{7},
B.~Zitzer\altaffilmark{9}
}

\altaffiltext{*}{Corresponding author: amandajw@iastate.edu}
\altaffiltext{1}{Department of Physics and Astronomy, Barnard College, Columbia University, NY 10027, USA}
\altaffiltext{2}{Physics Department, McGill University, Montreal, QC H3A 2T8, Canada}
\altaffiltext{3}{Department of Physics and Astronomy, University of California, Los Angeles, CA 90095, USA}
\altaffiltext{4}{Department of Physics, Washington University, St. Louis, MO 63130, USA}
\altaffiltext{5}{Fred Lawrence Whipple Observatory, Harvard-Smithsonian Center for Astrophysics, Amado, AZ 85645, USA}
\altaffiltext{6}{School of Physics, University College Dublin, Belfield, Dublin 4, Ireland}
\altaffiltext{7}{Santa Cruz Institute for Particle Physics and Department of Physics, University of California, Santa Cruz, CA 95064, USA}
\altaffiltext{8}{School of Physics and Astronomy, University of Leeds, Leeds, LS2 9JT, UK}
\altaffiltext{9}{Argonne National Laboratory, 9700 S. Cass Avenue, Argonne, IL 60439, USA}
\altaffiltext{10}{School of Physics, National University of Ireland Galway, University Road, Galway, Ireland}
\altaffiltext{11}{Astronomy Department, Adler Planetarium and Astronomy Museum, Chicago, IL 60605, USA}
\altaffiltext{12}{Department of Physics, Purdue University, West Lafayette, IN 47907, USA }
\altaffiltext{13}{Department of Physics, Grinnell College, Grinnell, IA 50112-1690, USA}
\altaffiltext{14}{School of Physics and Astronomy, University of Minnesota, Minneapolis, MN 55455, USA}
\altaffiltext{15}{Department of Astronomy and Astrophysics, University of Chicago, Chicago, IL, 60637}
\altaffiltext{16}{Department of Astronomy and Astrophysics, 525 Davey Lab, Pennsylvania State University, University Park, PA 16802, USA}
\altaffiltext{17}{DESY, Platanenallee 6, 15738 Zeuthen, Germany}
\altaffiltext{18}{Institute of Physics and Astronomy, University of Potsdam, 14476 Potsdam-Golm, Germany}
\altaffiltext{19}{Department of Physics and Astronomy, University of Utah, Salt Lake City, UT 84112, USA}
\altaffiltext{20}{Department of Physics and Astronomy, University of Iowa, Van Allen Hall, Iowa City, IA 52242, USA}
\altaffiltext{21}{Columbia Astrophysics Laboratory, Columbia University, New York, NY 10027, USA}
\altaffiltext{22}{Department of Physics and Astronomy and the Bartol Research Institute, University of Delaware, Newark, DE 19716, USA}
\altaffiltext{23}{Enrico Fermi Institute, University of Chicago, Chicago, IL 60637, USA}
\altaffiltext{24}{Physics Department, Columbia University, New York, NY 10027, USA}
\altaffiltext{25}{Department of Physics and Astronomy, DePauw University, Greencastle, IN 46135-0037, USA}
\altaffiltext{26}{Department of Physics and Astronomy, Iowa State University, Ames, IA 50011, USA}
\altaffiltext{27}{Saha Institute of Nuclear Physics, 1/AF Bidhannagar, Sector-1 Kolkata-700064, India}
\altaffiltext{28}{Kavli Institute for Cosmological Physics, University of Chicago, Chicago, IL 60637, USA}
\altaffiltext{29}{Department of Physics, Anderson University, 1100 East 5th Street, Anderson, IN 46012}
\altaffiltext{30}{Department of Life and Physical Sciences, Galway-Mayo Institute of Technology, Dublin Road, Galway, Ireland}
\altaffiltext{31}{School of Physics and Center for Relativistic Astrophysics, Georgia Institute of Technology, 837 State Street NW, Atlanta, GA 30332-0430}
\altaffiltext{32}{Department of Physics, Grand Valley State University, Allendale, MI 49401, USA}
\altaffiltext{33}{CRESST and Astroparticle Physics Laboratory NASA/GSFC, Greenbelt, MD 20771, USA.}
\altaffiltext{34}{University of Maryland, Baltimore County, 1000 Hilltop Circle, Baltimore, MD 21250, USA.}
\altaffiltext{35}{Physics Department, California Polytechnic State University, San Luis Obispo, CA 94307, USA}
\altaffiltext{36}{Department of Applied Physics and Instrumentation, Cork Institute of Technology, Bishopstown, Cork, Ireland}

\begin{abstract}

We report the discovery of an unidentified, extended source of very-high-energy
(VHE) gamma-ray emission, VER~J2019+407, within the radio shell of the
supernova remnant \snr\/, using 21.4 hours of data taken by the
\veritas\/ gamma-ray observatory in 2009.  These data confirm the preliminary
indications of gamma-ray emission previously seen in a two-year (2007-2009)
blind survey of the Cygnus region by \veritas\/.  VER~J2019+407, which is
detected at a post-trials significance of \posttrials\ in the 2009 data, is
localized to the northwestern rim of the remnant in a region of enhanced radio
and X-ray emission.  It has an intrinsic extent of \fittedextension\ and
its spectrum is well-characterized by a differential power law ($\rm dN/dE =
N_{0} \times (E/TeV)^{-\Gamma}$) with a photon index of \specindexg\ and a flux
normalization of \fluxnorm\/.  This yields an integral flux of \integralflux\ above
\threshold\/, corresponding to \fluxpercentage\ of the Crab Nebula flux.  We
consider the relationship of the TeV gamma-ray emission with the GeV gamma-ray emission seen
from \snr\/ as well as that seen from a nearby cocoon of freshly accelerated cosmic rays.  Multiple scenarios are considered as possible origins for the TeV gamma-ray emission, including hadronic particle acceleration at the supernova remnant shock.

\end{abstract}



\section{Introduction}\label{sec:intro}

The stereoscopic imaging atmospheric Cherenkov telescope array VERITAS
completed a two-year survey of the Cygnus region of the
Galactic Plane in very high energy (VHE) gamma rays \citep{VERITASCygnusFermi}.  The survey covered the
area between Galactic longitudes $67\,^{\circ}$ and $82\,^{\circ}$ and Galactic
latitudes $-1\,^{\circ}$ and $4\,^{\circ}$, a region chosen because it
contains both a high density of material and a significant population of
potential VHE gamma-ray emitters, including a number
of pulsar wind nebulae (PWNe) and supernova remnants (SNRs).  As acceleration of
either electrons or nuclei above 1 TeV within these stellar remnants can also
generate gamma-ray emission, detection of gamma rays from these objects may
provide insight into the nature of both the acceleration process within the
remnant and the nature of the accelerated population of particles \citep{1987PhR...154....1B,1991ApJ...382..242E,drury1994,Ellison2007}.  Relativistic electrons
can produce gamma rays via both non-thermal bremsstrahlung and inverse-Compton (IC) scattering off nearby
optical, IR, or microwave photons.  Protons and
heavier atomic nuclei generate gamma rays via the decay of neutral pions
produced by their interaction with interstellar material. Identifying gamma-ray
emission from SNRs that is produced by pion decay could provide insight into the origin of
Galactic cosmic rays, by providing evidence that SNRs are sites of
hadronic cosmic ray acceleration within the Galaxy.

While the initial survey yielded no clear source detections, the blind search suggested possible VHE gamma-ray emission at
several locations, most notably in the vicinity of the \snr\ ($\gamma$-Cygni).  SNR~G78.2+2.1 is a $\thicksim\!1^{\circ}$ diameter
supernova remnant (SNR) $\sim 1.7$~kpc  distant, with a shell-like radio and
X-ray structure (\citealt{Higgs1977,Lozinskaya2000}).  It is considered young to middle-aged at
$\sim\!7000$ years \citep{Higgs1977,Landecker1980,Lozinskaya2000} and in an early phase of adiabatic expansion into a medium of fairly low density
\citep{Lozinskaya2000}.  \citet{Gosachinskij2001} also identifies a slowly
expanding {\sc H{\thinspace}i} shell immediately surrounding the radio shell.
\citet{Lozinskaya2000} suggest this shell was created by the progenitor stellar
wind.

Most of the radio and X-ray emission lies in distinct northern and southern features \citep{Zhang1997,uchiyama}.
The northern region is characterized by enhanced thermal X-ray emission, suggestive of shocked gas \citep{uchiyama}, as well as
strong optical emission with sulfur lines \citep{Mavromatakis2003}.  However, it evinces little to no CO emission \citep{landp}.
The gamma-ray satellite \fermi\ sees diffuse gamma-ray emission above 10 GeV from the entire remnant \citep{Lande} and has
discovered a gamma-ray pulsar, \psr\/, at the center of the remnant \citep{fermipulsarcatalog,onefgl}.  This pulsar, which has a spectral cutoff of 3.0 GeV, also has a low luminosity ($1.1\times 10^{35}$ ~erg~s$^{-1}$) and a spin-down age (76.8~kyr)
much greater than the estimated age of \snr\/.  \citet{Trepl2010} argue, however, that {\psr} was likely born with something close to its current spin period, in which case its spin-down age is not indicative of its actual age.  It therefore remains plausible that \psr\ is the remnant of {\snr}'s progenitor star \citep{Trepl2010}.
Except where otherwise noted, all GeV and TeV sources are considered to be at a distance of 1.7 kpc, the estimated distance of \snr\ \citep{Higgs1977,Lozinskaya2000}.

Subsequent to the completion of the VERITAS Cygnus region survey,
follow-up observations were taken of the VHE gamma-ray source candidate near
\snr\/.  Based on those further observations, we report here the discovery
of an extended, unidentified source of VHE gamma rays that lies within an area
of enhanced radio emission along the northwestern shell of SNR G78.2+2.1.
The observational details are described in
\S\ref{sec:observations}; the analysis and results are presented in
\S\ref{sec:analysis} and \S\ref{sec:results}, respectively.  We discuss the
nature of the TeV gamma-ray emission in \S\ref{sec:discussion}.

\section{Observations}\label{sec:observations}

The \veritas\ Observatory consists of an array of four imaging
atmospheric Cherenkov telescopes, located at the Fred Lawrence Whipple
Observatory in southern Arizona \citep{Holder2008}.  Each
telescope is equipped with a $3.5^{\circ}$ field-of-view (FOV),
499-pixel photomultiplier tube camera.  The array is run in a coincident mode,
requiring a minimum of two out of four telescopes to trigger in order
for an event to be recorded. Data included in this paper were taken
with an upgraded array configuration that has improved sensitivity and gamma-ray
point-spread function (PSF) \citep{2009arXiv0912.3841P} relative to that used during the Cygnus region survey.
This configuration has $\sim\!15-25\%$ energy resolution between
100~GeV and 30 TeV and a $5\sigma$ (standard deviation above background) point source sensitivity of $1\%$ of
the Crab Nebula flux above 300~GeV in less than 30 hours of
observation at zenith angles less than $30^{\circ}$.

The observations under discussion were motivated by, but do not include, data
from the \veritas\ Cygnus region survey \citep{VERITASCygnusFermi}.  The
initial two-year survey covered the area of the remnant with an effective
exposure time of \surveyefflivetime\ and showed evidence of an extended
gamma-ray excess.
The approximate location of the survey excess was re-observed with a set of
dedicated pointings during September-November of 2009, for a total
live-time of \obslivetime\/. These observations were performed in ``wobble''
mode, with the center of the VERITAS field-of-view offset by $0.6^{\circ}$ from
the target position (RA $\rm 20^{h}19^{m}48^{s}$, Dec $\rm +40^{\circ}54'00''$).
The zenith angle was restricted to $10^{\circ}-30^{\circ}$ for these
observations.
Approximately $10\%$ of the data were taken at a time when three out of the four \veritas\ telescopes were operational; the remaining $90\%$ was taken with the full array.

\section{Analysis}\label{sec:analysis}

Images from these data are calibrated according to the standard procedure
described in \citet{Cogan2007} and cleaned using the procedure described in \citet{Daniel2007}.  After calibration, the primary
photon direction is calculated via stereoscopic reconstruction based on the
intersection of image primary axes.  To ensure that the primary photon
direction and energy can be well-reconstructed, an initial event selection is
applied, requiring that events have three images passing the following
criteria: more than four pixels per image, an image centroid no more than
$1.43^{\circ}$ from the camera center, and a total integrated charge per image
of at least \sizepe\ photoelectrons.


Calibrated images are described in terms of a second-moment parameterization \citep{hillas1985}.  Cosmic-ray
background is rejected using selection criteria applied to two composite parameters based on these moments:
mean-scaled length (MSL) and mean-scaled width (MSW) \citep{aharonianms}.  We impose the requirements {\mslrange} and
{\mswrange}; in addition, we require the angle between the reconstructed gamma-ray
arrival direction and the source position to be less than $0.23^{\circ}$.  The chosen background-rejection criteria are optimized for moderate-strength ($\sim 5\%$ of the Crab nebula flux) extended sources.  Together with the image quality requirements they impose an energy threshold for this measurement of \threshold.

To minimize the number of independent search elements, our search is restricted to a
pre-defined circular region with radius $0.25^{\circ}$ centered on the target position.  In the
imaging analysis and source morphology studies the ring background model \citep{Aharonian2005} is used to estimate the residual cosmic ray background; the reflected-region model \citep{Aharonian2001} is used when extracting the spectrum.  We also excluded from the
background estimation circular regions with radius $0.3^{\circ}$ around  four bright stars in the FOV ($\gamma$
Cygni, P Cygni, 40 Cygni, and HIP100069) as well as two overlapping circular
$0.4^{\circ}$-radius regions used to approximate the profile of the excess seen in the
\veritas\ survey data \citep{VERITASCygnusFermi,2011ICRC....7..161W}. All results reported here
have been verified by an independent calibration and analysis chain.

\section{Results}\label{sec:results}

\begin{figure}
\plotone{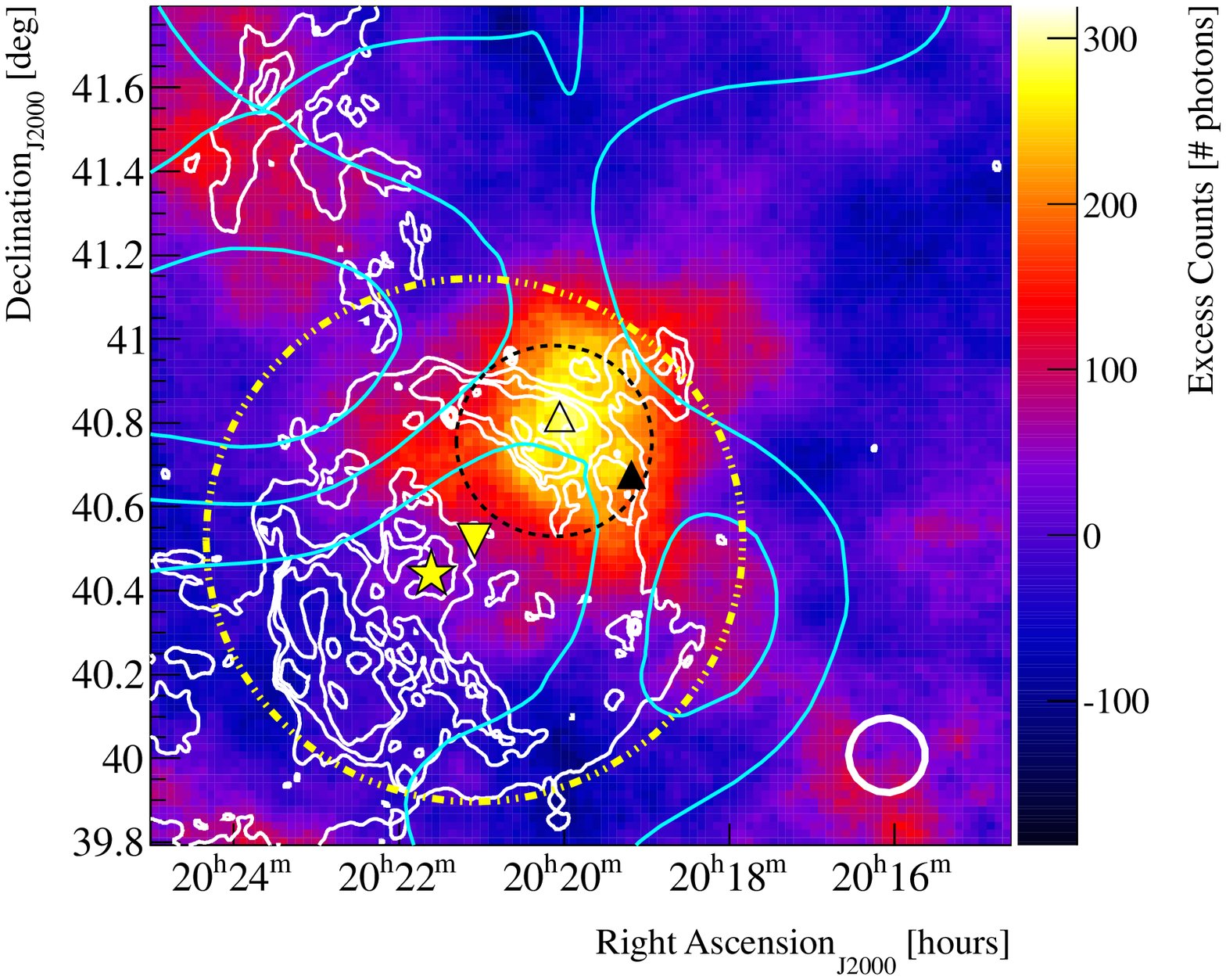}
\caption{Background-subtracted gamma-ray counts map of \snr\ showing the \veritas\ detection of \ver\ and its fitted extent (black dashed circle).    The supernova remnant is delineated by CGPS 1420 MHz continuum radio contours at brightness temperatures of 23.6K, 33.0K, 39.6K, 50K and 100K (white)
\citep{CGPS}; the star symbol shows the location of the
central gamma-ray pulsar \psr.  The inverted triangle and dot-dashed circle (yellow) show the fitted centroid and extent of the emission detected by \fermi\ above 10 GeV.
The open and filled triangles (black) show the positions of \fermi\ catalog sources \fgl\ and \fgltwo\ which have been subsumed into the extended GeV emission from the entire remnant.  The 0.16, 0.24, and 0.32 photons/bin contours of the \fermi\ detection of the Cygnus cocoon are shown in cyan.
The white circle (bottom right
corner) indicates the 68\% containment size of the \veritas\ gamma-ray PSF for this analysis.
\label{fig:excesswithoverlay} }
\end{figure}

\begin{figure}
\plotone{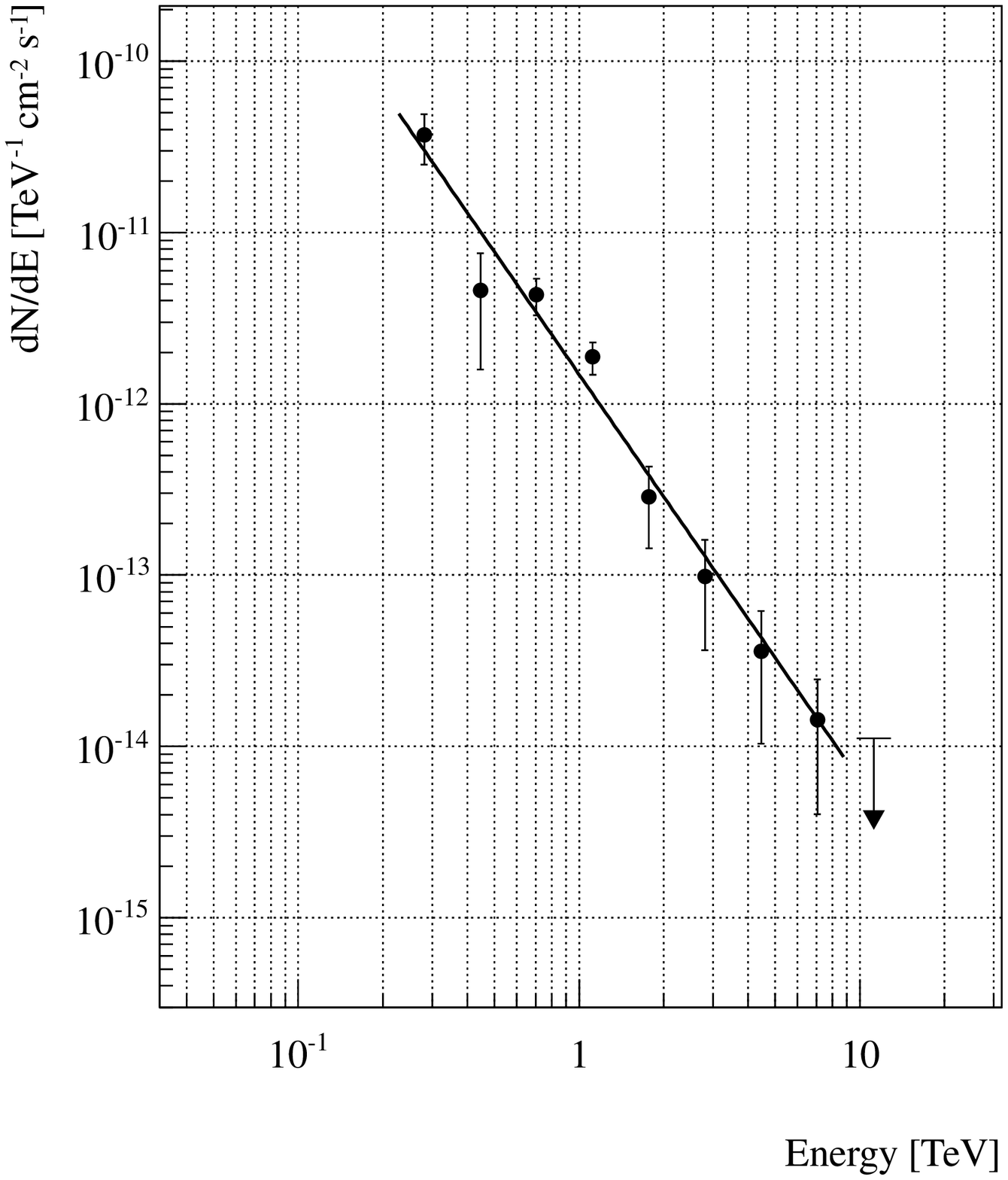}
\caption{Spectrum of
\ver, derived from four-telescope data only.  Points are the \veritas\ spectrum while the arrow indicates the upper limit on emission at 11 TeV.  The solid line shows a power-law fit with a spectral index of \specindexg\ and a flux normalization of \fluxnorm\/.\label{fig:tevspec}}
\end{figure}

Figure~\ref{fig:excesswithoverlay} displays the background-subtracted,
acceptance-corrected TeV image of the region of \snr\/.  A clear signal with
\excesscounts\ net counts is detected at the location of the northern rim of
the remnant.  This signal is significant at the \posttrialssig\ level after
accounting for all test points in the pre-defined $0.25^{\circ}$ search region.  Figure~\ref{fig:excesswithoverlay} also shows the locations of the gamma-ray pulsar \psr\ (\fglpsr), \centroidpulsaroffset\ from \ver\ at the center of the SNR, and the centroid
of the emission from the remnant seen by \fermi\ above 10 GeV.

The morphology of \ver\ is derived from a binned extended maximum-likelihood fit to the counts map before background subtraction.  The cosmic ray component is modeled as an exposure-modulated flat background and the source by a symmetric two-dimensional Gaussian convolved with the VERITAS point spread function (PSF) ($68\%$ containment radius of \containment, derived from an identically processed observation of the Crab Nebula).  We find a fitted extension of \fittedextension.  The fitted centroid coordinates are \fittedcentroidsexg ; however, we maintain the identifier \ver\ for the source, which was originally assigned on the basis of a preliminary centroid estimation.
The statistical uncertainty in this location is $0.03^{\circ}$, with a combined systematic uncertainty in the position, due to the telescope pointing error and systematic errors of the fit itself, of $0.018^{\circ}$.

Figure \ref{fig:tevspec} shows the spectrum derived from the
reconstructed gamma-ray events within \specradius\/ from the center
of the search region; runs where only three of four telescopes were operational have been excluded from this sample.  The threshold for the spectral analysis is \threshold\/ and the energy resolution is \energyres\/ at 1
TeV.  The photon spectrum is fit well (\spechisq\ ) by a differential power law in energy, $\rm dN/dE = N_{0}
\times (E/TeV)^{-\Gamma}$, between \analysisthreshold\/ and 10 TeV, with a photon index of
\specindex\/ and a flux normalization at 1 TeV of {\fluxnorm}. The
integral flux above \threshold\/ (\integralflux) corresponds
to \fluxpercentage\/ of the Crab Nebula flux above that energy.
No other significant TeV source is found in the maps.

\begin{figure}
\plotone{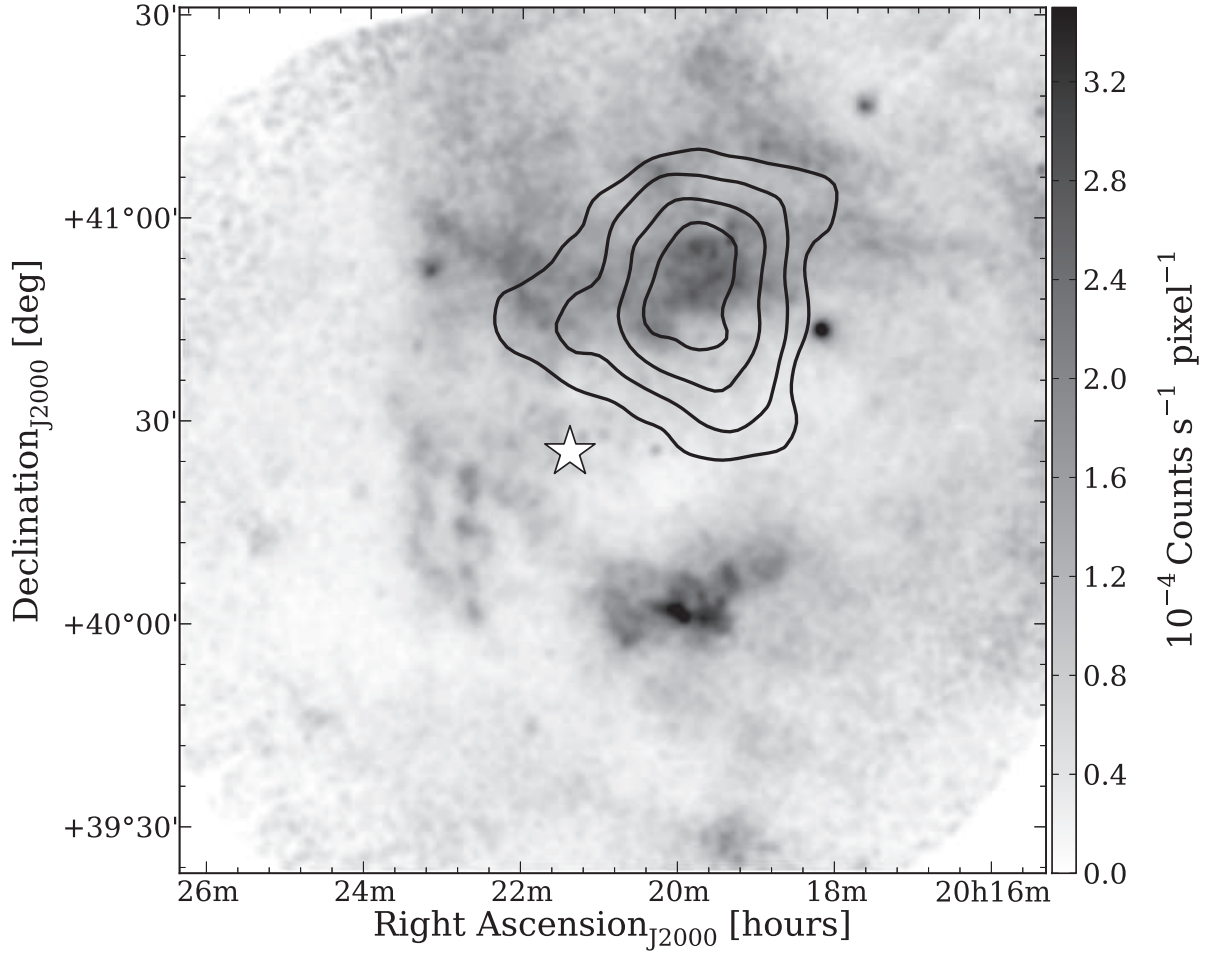}
\caption{ROSAT PSPC X-ray view of \snr\ between 1 and 2 keV.  The \ver\ smoothed photon excess contours (100, 150, 210 and 260 photons) are superimposed.  The image is composed of a mosaic of six exposure and vignette-corrected overlapping observations, smoothed using a $5 \times 5$ pixel boxcar filter. The lower energy bound was selected to reject the background flux from the Galactic Plane.  The location of the gamma-ray pulsar \psr\ is marked with a white star.}
\label{fig:rosat}
\end{figure}

\begin{figure}
\plotone{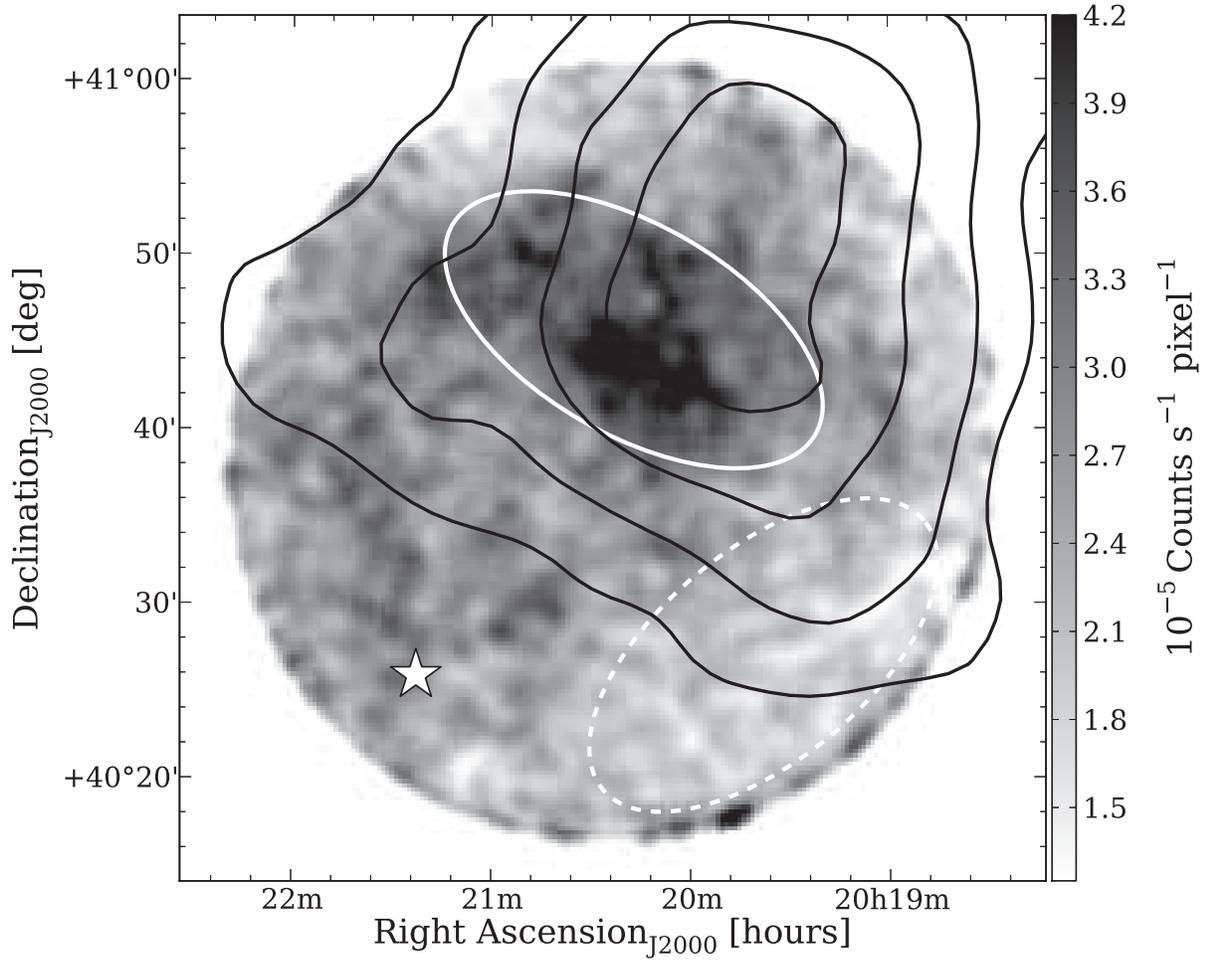}
\caption{ ASCA X-ray view of G78.2+2.1 between 1 and 3 keV, overlaid with the \ver\ smoothed photon excess contours (100, 150, 210 and 260 photons).  The region used to extract a spectrum and the corresponding background region are indicated by white solid and dashed ellipses, respectively.  A white star marks the position of \psr .}
\label{fig:asca}
\end{figure}

\begin{figure}
\plotone{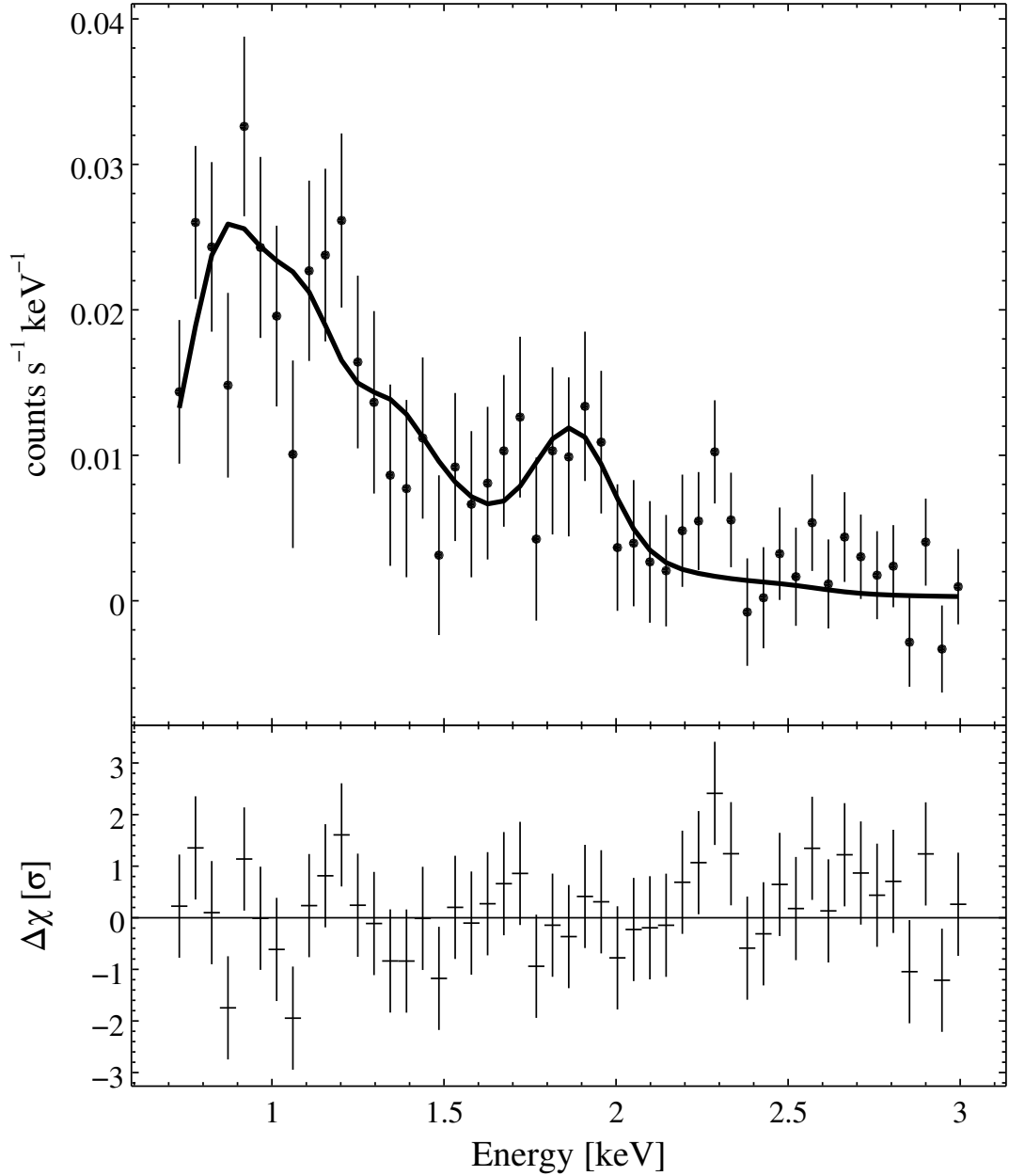}
\caption{
Top Panel --- ASCA X-ray spectrum of the region of enhanced X-ray emission coincident with \ver\/, as shown in Figure~\ref{fig:asca}.
The solid line shows the fit of a Raymond-Smith thermal plasma model with parameters as given in the text.  We identify the line at 1.9~keV as due to Si.
Bottom Panel --- $\Delta\chi$ residuals (residual divided by the statistical error) for the best-fit model.}
\label{fig:ascaspec}
\end{figure}

\subsection{X-Ray Observations}\label{sec:x-ray}

Figure~\ref{fig:rosat} presents an exposure-corrected ROSAT PSPC $1-2$~keV image of \snr\ generated from archival data (see \citet{1996MNRAS.281.1033B} for observational details).  The TeV gamma-ray emission overlaps a region of enhanced X-ray emission coincident with the bright radio arcs associated with the northern rim of the SNR shell.
As the highly absorbing column to the remnant screens out most source photons below $\sim 1$~keV, we excluded all photons below this energy to suppress background.
Several bright X-ray sources are associated with known stars and have been identified in previous work, but none of these overlap the VHE gamma-ray-emitting region \citep{Becker2004,Weisskopf2006}.

To better study the enhanced X-ray emission overlapping the TeV contours, we also re-analyzed the data from
ASCA Sequence \#25010000 previously studied by \citet{uchiyama}.  Figure~\ref{fig:asca} shows the $0.7-3.0$ keV exposure-corrected X-ray map, generated by co-adding data from the two gas imaging spectrometers.  The image is consistent with that seen by \rosat.  We extracted source and background spectra from the two $12^{\prime}\times 24^{\prime}$ elliptical regions displayed in Figure~\ref{fig:asca} and generated count-weighted response files appropriate for diffuse emission.
The source region, centered on coordinates RA~$20^{\rm h}$~$20^{\rm m}$~$17^{\rm s}$, Dec~$+40^{\circ}$~$45^{\prime}$~$41^{\prime\prime}$ (J2000) and oriented with position angle $60^{\circ}$, contains the bulk of the X-ray emission located within the \veritas\ contours.  The background ellipse, centered at RA~$20^{\rm h}$~$19^{\rm m}$~$38^{\rm s}$, Dec~$+40^{\circ}$~$27^{\prime}$~$02^{\prime\prime}$ (J2000) with position angle $130^{\circ}$, is placed where the X-ray emission is evidently at background level in the ROSAT image.    The source spectrum was grouped with a minimum of 20 counts per channel and fitted in the $0.7-3.0$~keV range using the XSPEC software package. The spectra are evidently soft with few net counts above $\sim 3$~keV after background subtraction.
A significant line feature is found at 1.9~ keV; we do not, however, find any evidence for the strong Ne line feature at $\simeq 0.9$ keV reported by \cite{uchiyama}.

We modeled the X-ray spectrum using an absorbed Raymond-Smith thermal plasma model (see Fig.~\ref{fig:ascaspec}).  This provides an adequate fit with $\chi^2/\rm DoF = 40.8/44$.  The best-fitted temperature is $kT = 0.57 \pm 0.14$ keV. Si is overabundant by a factor of $2.0^{+1.8}_{-1.2}$ relative to solar. The column density is $N_H = \rm (3.7 \pm 2.0) \times 10^{21} \thinspace cm^{-2}$ and the normalization is $\rm N=1.8\times 10^{-3} cm^{-5}$.  The
absorption-corrected flux for this model is found to be $6.0 \times 10^{-12} \rm \, erg \, cm^{-2} \, s^{-1}$ in the 0.5$-$8.0~keV band.  We also considered that the plasma behind a shock expanding in a low-density medium might not have had time to reach ionization equilibrium given the assumed SNR lifetime, in which case the spectrum would be better modeled by a non-equilibrium ionization model.  We attempted to fit the ASCA spectrum with an absorbed non-equilibrium ionization (NEI) model, the ``vgnei'' model of XSPEC, but were unable to obtain good constraints on the fitted parameters.

We also attempted to constrain the flux of any possible power-law component in the spectrum by adding a power-law with photon index fixed to 2.0 and then varying the power-law normalization until the $\chi^2$ increased by 2.7 ($90\%$ confidence level) relative to the best fitted model with no power-law component.  We place an upper limit on the flux of a power-law component of $1.9 \times 10^{-12} \rm \, erg \, cm^{-2} \, s^{-1}$ in the 0.5$-$8.0~keV band.

We note that our result differs significantly from that given in \cite{uchiyama}.  While we have chosen a source region similar to their ``R1'' source region, we find that our fits to the ASCA spectrum require neither an additional power-law component nor a large Ne IX line feature.  The discrepancy seems to be caused by the choice of background region.  We selected a background region that was as close as possible to the source, in order to account for the large variance in the Galactic emission in this part of the plane.  \cite{uchiyama} considered four nearby ASCA pointings and chose the one with the least source contamination, $\sim 3.5\arcdeg$ from the center of the ``R1'' region.  A choice of background region similar to Uchiyama's allows us to reproduce the previously-reported nonthermal component and spectral features.


\section{Discussion}\label{sec:discussion}

We have discovered a spatially extended source of VHE gamma-ray emission, \ver , located on the northwest rim of the shell-type supernova remnant \snr.  The TeV gamma-ray emission is coincident with the leading edge of a prominent arc-shaped radio continuum feature along the SNR rim (Fig.~\ref{fig:excesswithoverlay}) and with a region of enhanced X-ray emission that we identify as thermal in origin (Figs.~\ref{fig:rosat},~\ref{fig:asca}, and \ref{fig:ascaspec}).  \ver\ lies near the inner edge of a slowly expanding {\sc H{\thinspace}i} shell immediately surrounding the radio shell identified in 21~cm line observations \citep{Gosachinskij2001}; it also lies near a region of bright [{\sc S{\thinspace}ii}] optical line emission within the SNR that is identified as shock-heated gas based on the [{\sc S{\thinspace}ii}]/H$\alpha$ line ratio \citep{Mavromatakis2003}.  There are also two GeV sources associated with \snr.  One is a pulsar, \fglpsr, located at the center of the remnant \citep{fermipulsarcatalog,onefgl}.  The other is an extended source candidate reported above 10 GeV by  \fermi\ \citep{Lande}.  A point source co-located with \ver\ was previously reported in the first and second \fermi\ catalogs but \citet{Lande} conclude that it was an artifact of failing to model the extended source above 10 GeV rather than a separate source.  Between G78.2+2.1 and the Cyg OB2 association (roughly $2.4^{\circ}$ away) lies a 50 parsec-wide area of 1-100 GeV gamma-ray emission, detected by \fermi\/, that has been interpreted as a cocoon of freshly-accelerated cosmic rays \citep{2011Sci...334.1103A}.

The TeV gamma-ray emission and the features observed in the X-ray, optical, and radio continuum can be most simply
related to the presence of shocks at the interaction of the supernova ejecta and the surrounding medium.
\snr\ has a relatively low interior density and, as previously noted, appears to be surrounded by a dense {\sc H{\thinspace}i} shell \citep{Gosachinskij2001}.
\citet{Lozinskaya2000} suggest
this {\sc H{\thinspace}i} shell was created by the progenitor stellar wind as it swept up the
ambient medium.
The interaction of the supernova shock with the {\sc H{\thinspace}i} shell might drive a shock into the dense shell and a reflected shock back into the cavity.
Emission arising from these shocks would then account for the observed [{\sc S{\thinspace}ii}] lines, enhanced thermal X-rays, and strong radio continuum emission near \ver\/.

Shocks can produce TeV gamma-ray emission via either inverse-Compton scattering of accelerated electrons or hadronic interactions of accelerated nuclei.
High-energy electrons capable of producing TeV photons via inverse-Compton scattering should also produce X-ray synchrotron radiation detectable
as a non-thermal power-law in the X-ray spectrum.
While our analysis of the ASCA X-ray spectrum shows no evidence for a non-thermal component, the upper limit on this component is not yet strong enough to rule out inverse-Compton scattering as the source of the TeV, if not the GeV, gamma-ray emission.
Deeper X-ray observations of \ver\ should provide better constraints.

The flux of TeV photons from hadronic processes depends on the energy available for shock acceleration and the density of the target material.  One of the prerequisites of this scenario is that the necessary density of target material, as inferred from the gamma-ray flux, be consistent with that estimated from measurements at other wavelengths.
Equation 9 in \cite{drury1994} gives (on the assumption that the spectrum of the charged particles is a power law with index of $-2$) the expected gamma-ray
flux above a given threshold energy as a function of the threshold energy $E$, the fraction $\theta$ of the SN kinetic energy $E_{SN}$ converted to cosmic rays,
the distance $d$ of the SNR, and the density $n$ of the target material,

\begin{equation}\label{eqn:DAV49} \rm{
    F(>E) \approx 9 \times 10^{-11} \ \theta \ (\frac{E}{1 \ TeV})^{-1.1} \ (\frac{E_{SN}}{10^{51} \ erg}}) \
    (\frac{d}{1 \ kpc})^{-2} \ (\frac{n}{1\ cm^{-3}})\ cm^{-2} \ s^{-1}.
\end{equation}

For $E = 320$~GeV, $d = 1.7$~kpc, $\theta \sim \rm{0.1}$ as expected for a remnant in the early Sedov phase, and assuming $E_{SN} = 10^{51}$~erg, which is consistent with estimates derived from optical data \citep{Mavromatakis2003}, we find that a
density of \initialdensity\ is required to produce the observed TeV flux.  However, this equation is for the total flux integrated over the whole spherical shell
of a SNR.  The gamma-ray emission from VER J2019+407 is confined to a relatively small portion of the SNR shell.  Thus, it is likely that only a fraction of the
SN blast wave participates in the shock acceleration producing the TeV gamma-ray emission.  Assuming an isotropic explosion, we account for this by scaling $E_{SN}$ by the ratio of the surface area covered by VER J2019+407 to the total surface area of the remnant.  Using the fitted 95$\%$ (68$\%$) containment radius of the TeV gamma-ray emission
gives a correction factor of 0.5 (0.14).  Once uncertainties in the source extension, integral flux, and SNR distance are also taken into account, the average density required to produce the observed TeV flux above \threshold\ increases to \inferredensities.

We find these densities consistent with our knowledge of the region from other wavelengths.
Given that no significant CO emission is detected in this area of the remnant, any target material must be atomic rather than molecular, with the necessary high density perhaps arising in the swept-up {\sc H i} shell.
\citet{Gosachinskij2001} estimated a gas density of 2.5~cm$^{-3}$ in the {\sc H i} shell surrounding the remnant, consistent with the range of densities derived from the TeV data.
Optical data also shed some light on densities in the vicinity of VER J2019+407.  Mavromatakis gives the ratio of [{\sc S{\thinspace}ii}] $\lambda$6716 to [{\sc S{\thinspace}ii}] $\lambda$6731 line fluxes at multiple locations within G78.2+2.1, one of which is close to (although not precisely coincident with) VER J2019+407.  All
of these ratios are close to one, and, within a plausible range of temperatures, imply post-shock densities on the order of a few 100~cm$^{-3}$.  Mavromatakis infers these densities to be arising behind a radiative shock and
estimates a primary shock velocity of $\sim 750$ km/s, consistent with that inferred from an X-ray temperature of $0.57$ keV.  He deduces pre-shock cloud densities on the order of $\sim 20$~cm$^{-3}$, which are not unreasonable given the average densities inferred here from the gamma-ray flux.  The shock velocities inferred from the optical and X-ray data, however, are too low for the forward shock to accelerate particles to TeV energies at the present time.  Any hadronic TeV gamma-ray emission would have to arise from particles accelerated during an earlier epoch that are only now interacting with the shell.

The limited angular extent of the TeV gamma-ray emission also raises questions.
If we view the SN blastwave as expanding in the low-density bubble blown by the progenitor stellar wind until it reaches the much higher-density cavity wall, why should the interaction favor one portion of the
cavity wall over the others?  Asymmetries in the shock propagation, variations in the density of the cavity wall, or the former presence of a cloud
(now evaporated by the shock) within the low density region could provide an explanation, as could the strength and orientation of the ambient magnetic field relative to the local shock velocity \citep{Ellison2007}.  Given the more symmetric and diffuse GeV gamma-ray emission seen from the remnant, asymmetric diffusion of accelerated particles out of the remnant might also play a role, as the highest-energy particles would be the first to escape.
However, at this stage there is no clear evidence for or against any of these possibilities.

In these scenarios, \snr\ could be responsible for some or all of the freshly-accelerated cosmic rays within the cocoon detected by \fermi\/.  However, the results shown here cannot be used to draw strong conclusions about the cocoon's relationship to \snr\ or to set a meaningful upper limit on cocoon emission above 300 GeV.
The cocoon contours from \citet{2011Sci...334.1103A} were derived from an analysis where \fgl\/, which is no longer considered an independent source, was included as part of the background model.  Since this would artificially reduce the intensity of the GeV gamma-ray emission seen near \ver\/, the \citet{2011Sci...334.1103A} contours provide a spatial reference only and should not be used to judge the relationship of the cocoon to \ver\/.
Moreover, gamma-ray emission on the scale of the cocoon (roughly four square degrees) cannot be detected with VERITAS using the ring-background estimation method, which will cause the source to self-subtract.  Analysis techniques better adapted to highly extended sources, combined with further data taken in alternative observation modes, will be required to confirm the presence or absence of cocoon emission and make a definitive statement about \snr\/'s role in feeding the cocoon.

\ver\ bears some similarity to other well known TeV sources.  H.E.S.S.\ observations of the shell-type SNR RCW 86 have revealed the presence of VHE
gamma rays suggestive of a shell-type morphology \citep{felix09}.  \citet{bpv09} have explained the TeV gamma-ray emission from RX J0852.0-4622 as arising
from hadronic emission in a wind-bubble scenario, similar in many aspects to the scenario proposed here.

Other explanations of \ver\ are also possible.  Nearly half of Galactic TeV
sources are identified with pulsar wind nebulae (PWNe).  TeV PWNe are typically
extended sources energized by electrons accelerated in the pulsar wind. Because
the lifetimes of the electrons can be long, the electrons can diffuse over
large distances from the pulsar and the centroid of the TeV gamma-ray emission is often
offset from the pulsar \citep{Hessels2008}. On the one hand, the luminosity of
\ver\ in the 1$-$10~TeV band---$2.5 \times 10^{33} \rm \, erg \, s^{-1}$
assuming a distance of 1.7~kpc---is 2\% of the spin-down power of \psr, a value
within the range seen for PWNe.  On the other, \citet{Hessels2008} examined the
offsets between TeV gamma-ray emission and pulsar location for 21 TeV sources, and the
pulsar, while offset from the centroid of the TeV gamma-ray emission, generally lies well
within the angular extent of the TeV source. In the single case for which the
angular offset is significantly larger than the angular size of the TeV
source---PSR J1702-4128/HESS J1702-420-- inspection of the TeV image shows TeV
emission at the pulsar location \citep{HESSUNID}.  By contrast, the offset
between \psr\ and \ver\ is several times larger than the angular size of \ver,
and we detect no TeV gamma-ray emission either at the location of \psr\ or in the
intervening region between \psr\ and the TeV source.

Finally, it is possible that \ver\ is the PWN of an unknown pulsar in the line-of-sight toward \snr.  This scenario could explain the GeV and radio continuum emission, but would ascribe the location of \ver\ near the [{\sc S{\thinspace}ii}] line emission and the enhanced thermal X-ray emission within \snr\ to chance superposition.
Such a scenario cannot be excluded at the current time, but could be tested by a sensitive search for the PWN in the X-ray band and also by radio or X-ray searches for the putative pulsar.

In summary, we have detected gamma-ray emission from a region of enhanced
radio, optical line, and X-ray emission in the northwestern shell of \snr. The
extended TeV source overlaps with GeV gamma-ray emission from the remnant but is notably offset
from the gamma-ray pulsar \psr\, which lies outside the 99\% confidence contour of
VER~J2019+407.  It seems most probable that VER J2019+407 arises from particles
(either hadronic or leptonic) accelerated within the SNR shock, although we
cannot yet rule out the possibility of a line-of-sight coincidence between the
remnant and an un-associated pulsar wind nebula. Deeper high energy
observations are needed to better constrain possible emission models.

This research is supported by grants from the U.S. Department of Energy Office of Science, the U.S. National Science Foundation and the Smithsonian Institution, by NSERC in Canada, by Science Foundation Ireland (SFI 10/RFP/AST2748) and by the Science and Technology Facilities Council in the U.K.  We acknowledge the excellent work of the technical support staff at the Fred Lawrence Whipple Observatory and at the collaborating institutions in the construction and operation of the instrument.  Dr. Weinstein and Dr Dwarkadas' research was also supported in part by NASA grant NNX11A086G.


\begin{thebibliography}{99}



\bibitem[Abdo et al.(2010a)]{fermipulsarcatalog} Abdo, A.~A., Ackermann, M., Ajello, M., et al.\ 2010a, \apjs, 187, 460

\bibitem[Abdo et al.(2010b)]{onefgl} Abdo, A.~A., Ackermann, M., Ajello, M., et al.\ 2010b, \apjs, 188, 405

\bibitem[Ackermann et al.(2011)]{2011Sci...334.1103A} Ackermann, M., Ajello, M., Allafort, A., et al.\ 2011, Science, 334, 1103

\bibitem[Aharonian et al.(2005)]{Aharonian2005} Aharonian, F., Akhperjanian, A. G., Aye, K.-M., et al. 2005, A\&A, 430, 865

\bibitem[Aharonian et al.(2008)]{HESSUNID} Aharonian, F., Akhperjanian, A. G., Barres de Almeida, U., et al.\ 2008, \aap, 477, 353

\bibitem[Aharonian et al.~(2009)]{felix09} Aharonian, F., Akhperjanian, A. G., Barres de Almeida, U., et al. 2009, \apj, 692, 1500

\bibitem[Aharonian et al.(2001)]{Aharonian2001} Aharonian, F., Akhperjanian, A., Barrio, J., et al. 2001, A\&A, 370, 112

\bibitem[Aharonian et al.(1997)]{aharonianms} Aharonian, F.A., Hofmann, W., Konopelko, A.K., \& V\"{o}lk, H.J. 1997, Astroparticle Physics, 6, 343

\bibitem[Becker et al.(2004)]{Becker2004} Becker, W., Weisskopf, M., Arzoumanian, Z., et al.\ 2004, \apj, 615, 897

\bibitem[Berezhko et al.(2009)]{bpv09} Berezhko, E. G., Puhlhofer, G., \& V\"{o}lk, H. J. 2009, A\&A, 505, 641

\bibitem[Blandford
\& Eichler(1987)]{1987PhR...154....1B} Blandford, R., \& Eichler, D.\ 1987, \physrep, 154, 1

\bibitem[Brazier et al.(1996)]{1996MNRAS.281.1033B} Brazier, K.~T.~S., Kanbach, G., Carrami{\~n}ana, A., Guichard, J.,
\& Merck, M.\ 1996, \mnras, 281, 1033


\bibitem[Cogan et~al.(2007)]{Cogan2007} Cogan, P., Acciari, V.~A., Amini, R., et al. 2007, \newblock in Proceedings of the 30th International Cosmic Ray Conference, ed. Rogelio Caballero, Juan Carlo D'Olivo, Gustavo Median-Tanco, Lukas Nellen, Federico A. S{\'a}nchez, \& Jos{\'e} F. Vald{\'e}s-Galicia, 3, 1385

\bibitem[Daniel et al.(2007)]{Daniel2007}
Daniel, M.~K., Acciari, V.~A., Amini, R., et al.  2007,
\newblock in Proceedings of the 30th International Cosmic Ray Conference, ed. Rogelio Caballero, Juan Carlo D'Olivo, Gustavo Median-Tanco, Lukas Nellen, Federico A. S{\'a}nchez, \& Jos{\'e} F. Vald{\'e}s-Galicia, 3, 1325

\bibitem[Drury et al.(1994)]{drury1994} Drury, L. O'C., Aharonian, F.A., and V\"{o}lk, H.J.
1994,  \aap, 287, 959-971

\bibitem[Ellison et al.(2007)]{Ellison2007} Ellison, D.C., Patnaude, D.J., Slane, P., Blasi, P., Gabici, S.\ 2007, \apj, 661, 879

\bibitem[Ellison
\& Reynolds(1991)]{1991ApJ...382..242E} Ellison, D.~C., \& Reynolds, S.~P.\ 1991, \apj, 382, 242


\bibitem[Gosachinskij(2001)]{Gosachinskij2001} Gosachinskij, I.~V.\
2001, Astronomy Letters, 27, 233

\bibitem[Hessels et al.(2008)]{Hessels2008} Hessels, J.~W.~T.,  Nice, D.~J., Gaensler, B.~M., et al.\ 2008, \apjl, 682, L41

\bibitem[Higgs et al.(1977)]{Higgs1977} Higgs, L.~A., Landecker,
T.~L., \& Roger, R.~S.\ 1977, \aj, 82, 718

\bibitem[Hillas(1985)]{hillas1985} Hillas, A. M. 1985,
\newblock in Proceedings of the 19th International Cosmic Ray Conference (La Jolla), ed. F.C. Jones,
3, 445

\bibitem[Holder et al.(2008)]{Holder2008}
Holder, J., Acciari, V.~A., Aliu, E., et al. 2008,
\newblock in American Institute of Physics Conference Series, Proceedings of the 4th International Meeting on High Energy Gamma-Ray Astronomy, ed. F. A. Aharonian, W. Hofmann, \& F. Rieger, 1085, 657

\bibitem[Ladouceur \& Pineault(2008)]{landp} Ladouceur, Y., \& Pineault, S.\ 2008,
\aap, 490, 197

\bibitem[Lande et al.(2012)]{Lande} Lande, J., Ackermann, M.,
Allafort, A., et al.\ 2012, \apj, 756, 5

\bibitem[Landecker et al.(1980)]{Landecker1980} Landecker, T.~L., Roger, R.~S., \& Higgs,
L.~A.\ 1980, \aaps, 39, 133

\bibitem[Lozinskaya et al.(2000)]{Lozinskaya2000} Lozinskaya, T.~A.,
Pravdikova, V.~V., \& Finoguenov, A.~V.\ 2000, Astronomy Letters,
26, 77

\bibitem[Mavromatakis(2003)]{Mavromatakis2003} Mavromatakis, F.\ 2003, \aap, 408, 237

\bibitem[Perkins et al.(2009)]{2009arXiv0912.3841P} Perkins, J.~S., Maier,
G., Aliu, E., et al. 2009, in 2009 \fermi\ Symposium, Washington,
D.C., 2009, ed. W. N. Johnson and D. J. Thompson, eConf
C0911022, arXiv:0912.3841

\bibitem[Taylor et al.(2003)]{CGPS} Taylor, A.~R., Gibson, S.~J., Peracaula, M., et al. 2003, \aj, 125, 3145

\bibitem[Trepl et al.(2010)]{Trepl2010} Trepl, L.,  Hui, C.~Y., Cheng, K.~S., et al.\ 2010, \mnras, 405, 1339-1348

\bibitem[Uchiyama et al.(2002)]{uchiyama} Uchiyama, Y., Takahashi, T., Aharonian, F.~A., \& Mattox, J.~R.\ 2002, \apj, 571, 866


\bibitem[Weinstein et al.(2011)]{2011ICRC....7..161W} Weinstein, A., Aliu, E., Arlen, T., et al. 2011, in Proceedings of the 32nd International Cosmic Ray Conference, 7, 161.

\bibitem[Weinstein et al.(2009)]{VERITASCygnusFermi}
Weinstein, A.,  Aliu, E., Arlen, T., et al. 2009, in \fermi\ Symposium, Washington,
D.C., 2009, ed. W. N. Johnson and D. J. Thompson, eConf
C0911022, arxiv:0912.4492

\bibitem[Weisskopf et al.(2006)]{Weisskopf2006} Weisskopf, M.C., Swartz, D.~A.; Carrami\~{n}ana, A.,\ et al.\ 2006, \apj, 652, 387

\bibitem[Zhang et al.(1997)]{Zhang1997} Zhang, X., Zheng, Y., Landecker, T.~L., \&
Higgs, L.~A.\ 1997, \aap, 324, 641

\end{thebibliography}
\end{document}